\documentclass[12pt,preprint]{aastex}

\shorttitle{Magnetic Flux Loss and Transport}
\shortauthors{Kubo et al.}

\begin{document}

\title{Magnetic Flux Loss and Flux Transport in a Decaying
Active Region}

\author{M. Kubo\altaffilmark{1}, B. W. Lites\altaffilmark{1},
T. Shimizu\altaffilmark{2}, \and K. Ichimoto\altaffilmark{3,4}}
\email{kubo@ucar.edu}

\altaffiltext{1}{High Altitude Observatory, National Center for
Atmospheric Research, P.O. Box 3000, Boulder, CO 80307. The National Center for
Atmospheric Research is sponsored by the National Science Foundation.} 
\altaffiltext{2}{Institute of Space and Astronautical Science, JAXA,
Sagamihara, Kanagawa 229-8510, Japan.} 
\altaffiltext{3}{National Astronomical Observatory of Japan, 2-21-1
Osawa, Mitaka, Tokyo 181-8588, Japan.} 
\altaffiltext{4}{Current address: Hida Observatory, Kyoto University,
Takayama, Gifu 506-1314, Japan.}

\begin{abstract}
We estimate the temporal change of magnetic flux perpendicular to the
 solar surface in a decaying active region by using a time series
 of the spatial distribution of vector magnetic fields in the photosphere. 
The vector magnetic fields are derived from full
 spectropolarimetric measurements with the Solar Optical Telescope 
 aboard \textit{Hinode}.
We compare a magnetic flux loss rate to a flux transport rate in a
 decaying sunspot and its surrounding moat region. 
The amount of magnetic flux that decreases in the sunspot and moat region
 is very similar to magnetic flux transported to the outer boundary of
 the moat region.
The flux loss rates [$(dF/dt)_{loss}$] of magnetic elements with
 positive and negative polarities are balanced each other around the
 outer boundary of the moat region. 
These results suggest that most of the magnetic flux in the sunspot is
 transported to the outer boundary of the moat region as moving
 magnetic features, and then removed from the photosphere by flux
 cancellation around the outer boundary of the moat region. 
\end{abstract}


\section{INTRODUCTION}
How and where is magnetic flux in sunspots removed from the photosphere?
Sunspot umbrae are sometimes split by formation of a light bridge, which
is a bright lane crossing the umbra \citep{Bray1964}.
Small magnetic features with a typical size less than 2$\arcsec$
called moving magnetic features \citep[MMFs;][]{Harvey1973} are generally
observed in the moat region that surrounds a decaying, mature sunspot. 
It has been reported that MMFs appear not only in the decaying phase
of sunspots but also in the growing phase \citep[e.g.][]{Wang1991}.
The MMFs mostly appear around the outer boundary of the sunspot, moving 
almost radially outward during their lifetime ranging from a few minutes
to 10 hr \citep{Harvey1973, Zhang2003, Hagenarr2005}. 
The formation of the light bridges and MMFs is closely related to
the fragmentation and disintegration of the sunspot magnetic flux.
Indeed, it has been observed that the net flux carried away from the
sunspot by MMFs is larger than the flux decrease in the sunspot
\citep{Martinez2002, Kubo2007a}.
This indicates that MMFs can be responsible for the flux loss of the
sunspot.

The mutual apparent loss of magnetic flux is often observed in the
line-of-sight magnetograms when one magnetic polarity element collides
with another polarity magnetic element in the photosphere.
This apparent flux loss is called ``magnetic flux cancellation'' as a
descriptive term.
It is observed that moving magnetic features often collide with 
apparently static opposite polarity magnetic features around the outer 
boundary of the moat region \citep{Martin1985, Yurchyshyn2001, Chae2004, 
Bellot2005} and widely believed that understanding the flux cancellation 
process around the moat boundary is the key to understand the dissipation 
of the sunspot flux from the photosphere. 
Three models have been proposed by \citet{Zwaan1987} to describe the flux
cancellation: (1) retraction of magnetic fields that connect an
emerged bipole, (2) submergence of $\Omega$-loop formed by magnetic
reconnection between the canceling two bipoles above the photosphere,
and (3) emergence of U-loop due to reconnection below the photosphere.
As expected in these models, horizontal magnetic fields are formed between
canceling magnetic elements \citep{Wang1993, Chae2004, Kubo2007b}.
However, whether upward or downward motions are observed in the
cancellation sites depends on the events and the positions in the
cancellation sites \citep{Harvey1999, Kubo2007b}. 
Therefore, the nature of the physical process driving magnetic flux 
cancellation is still an open issue.

This study attempts to address a basic question how much magnetic flux
is carried away from the sunspot to the outer boundary of the moat
region and is subsequently removed from the photosphere.
Because it has been difficult to measure the magnetic field vector under
stable seeing conditions for the period longer than a typical lifetime of 
MMFs, the flux loss rate of the sunspot, flux transport rate due to MMFs, 
and flux cancellation rate have been independently estimated by using 
different data sets. 
A time series of spectropolarimetric measurements with the Solar Optical
Telescope \citep[SOT;][]{Tsuneta2008} aboard the \textit{Hinode} satellite
\citep{Kosugi2007} allows us, for the first time, to estimate an accurate
flux change without any effects of atmospheric seeing.
Moreover, the high spatial resolution observations with the SOT decrease
the likelihood of spurious magnetic cancellation events, i.e., those for 
which magnetic elements with opposite polarities are located entirely 
within a resolution element and will dramatically increase the reliability 
of the results presented.

\section{OBSERVATIONS}
\begin{figure}
\epsscale{1.0}
\plotone{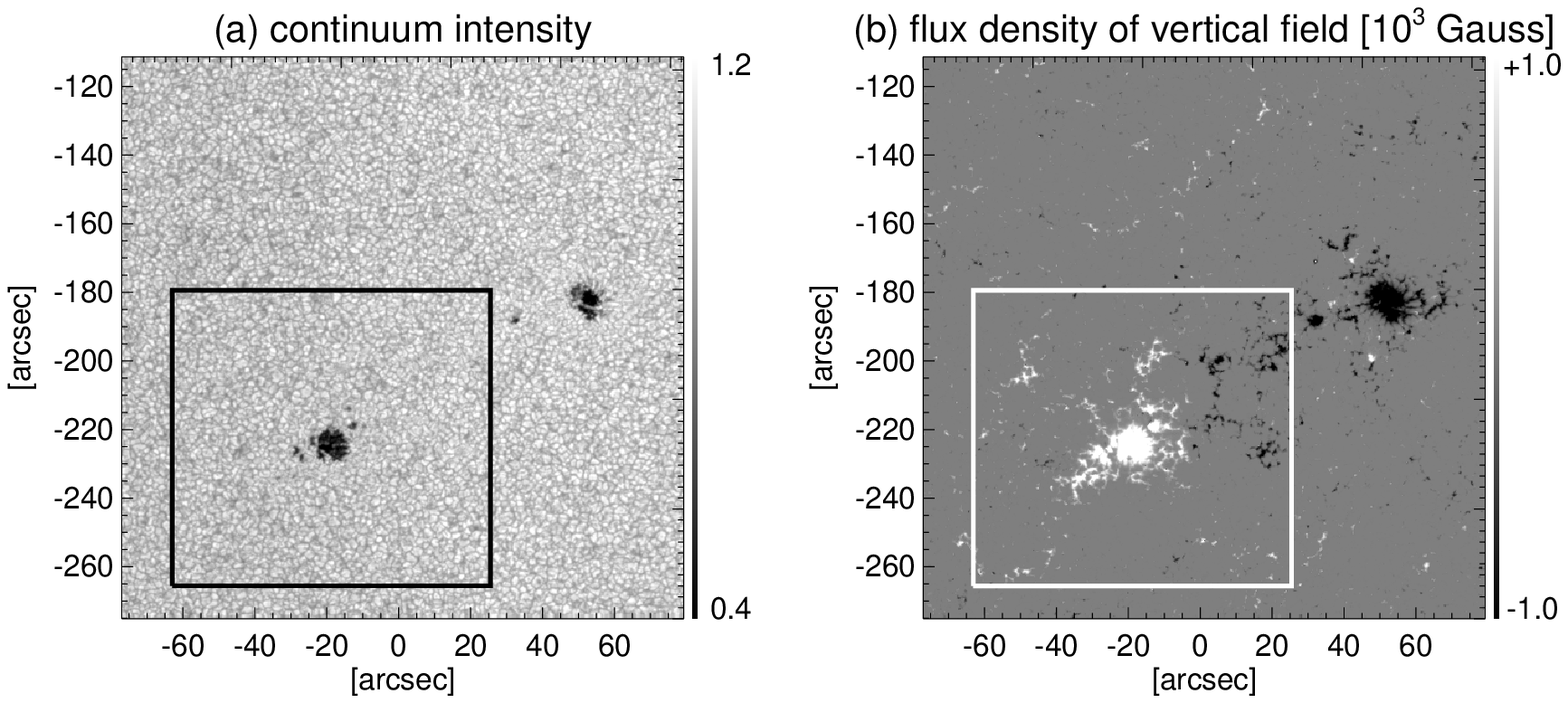}
\caption{(\textit{a}) Continuum intensity and (\textit{b}) magnetic flux
 density of vertical field in NOAA AR 10972. 
The continuum intensity is normalized to the mean intensity of the
 quiet area outside the sunspots.
These panels are made from observations with the \textit{Hinode} SP from
 15:05 to 15:37 on 2007 October 6. 
The box is identical to the field of view of Fig.~\ref{flux_image_evo}.
The positions in the vertical and horizontal axes are given with respect
 to the center of the solar disk.
} 
\label{ar10972}
\end{figure}

NOAA AR 10972 emerged on 2007 October 5 and formed a small bipolar sunspot, 
as shown in Figure~\ref{ar10972}. 
Both the leading and following sunspots completely disappeared at the end 
of October 8, leaving only  plage regions. 
The \textit{Hinode} SOT started to observe this active region from 14:00
on October 6, so that we did not observe the growing phase of the active
region. 
However, SOT provided a good data set for the decay phase.
We selected observations of the following sunspot from 15:05 on October
6 to 07:12 on October 8, at which time the sunspot had significantly decayed
(Fig.~\ref{flux_image_evo}).
The reason why we focused on only the following sunspot was that a part of
moat region surrounding the leading sunspot was outside the field of view.
In this period, the spectropolarimeter (SP) of the SOT scanned the active
region with the field of view of $152\arcsec \times 164\arcsec$ every
1-2 hr, except for two 4 hr gaps.
This provided us a time series of spatial distributions of the full
polarization state for two photospheric Fe {\footnotesize I} lines at
6301.5 {\AA} and 6302.5 {\AA}.
The slit scanning step was 0.297$\arcsec$ with an integration time of
3.2 seconds (Fast mapping mode).
SP took about 32 minutes to complete each scan.
The pixel scale along the slit was 0.320$\arcsec$, which was binned 2
pixels with the original spatial resolution of SP.
In the same period, the narrowband filter imager (NFI) of the SOT
continously obtained Stokes \textit{I} and \textit{V} images of the
active region at a wavelength of -172 m{\AA} from the center of the lower
chromospheric Na {\footnotesize D} line at 5896 {\AA} with a 2 minute cadence. 
The field of view was $276\arcsec \times 164\arcsec$ with a pixel
sampling of 0.16$\arcsec$.

\section{DATA ANALYSIS}
\subsection{Magnetic Flux Density}
The temporal change of magnetic flux in and around the following sunspot
was estimated from observations with the SP.
The Stokes profiles of the two Fe lines were calibrated with a standard
routine (``SP$_{-}$PREP''; B. W. Lites et al. 2008, in preparation).
The magnetic field vector and thermodynamic parameters were derived from 
the calibrated Stokes profiles with a Milne-Eddington Stokes inversion
(T. Yokoyama et al. 2008, in preparation).
The inversion code provided us magnetic field vector in a line-of-sight
frame.
A two-component model atmosphere, in which the photospheric atmosphere
was composed of a magnetized atmosphere (polarized light) and a
non-magnetized atmosphere (non-polarized light), was assumed in the
inversion code. 
We estimated a flux density of magnetic field vertical to the solar
surface for the pixels that have the degree of polarization larger than
0.5 $\%$ as 
\begin{equation}
F = \frac{1}{\cos\theta}f|{\bf{B}}|\cos\gamma,\label{eq_flux_d}
\end{equation}
where a heliocentric angle ($\theta$), field strength ($|{\bf{B}}|$),
inclination angle ($\gamma$) with respect to the local vertical, and
filling factor ($f$).
The filling factor is an areal percentage of the magnetized atmosphere
in each pixel.

A disambiguation of azimuth angles was needed to convert from the
inclination with respect to the line-of-sight direction into the
inclination with respect to the local vertical.
To perform the disambiguation, we selected the azimuth angles closer 
to the azimuth of potential fields at the photospheric surface, and
then interactively determined the azimuth to reduce discontinuities 
of azimuth and inclination angles by using the AZAM utility
\citep[written in IDL by P. Seagraves;][]{Lites1995}.
The active region has a simple bipolar structure with simple unipolar
spots.
For such cases the ambiguity resolution with the AZAM is almost
successfully performed \citep{Metcalf2006}.
Furthermore, the sunspot was located near the disk center and the
heliocentric angle ranges from 11$\degr$ to 25$\degr$. 
This means that projection effects of errors in magnetic field
inclination due to the disambiguation of azimuth angle are small. 

We applied an image cross-correlation for the magnetic flux density maps
in order to align the SP maps obtained at different times.
The area around the following sunspot was used in the image
cross-correlation in order to accurately remove the proper motion of the
following sunspot. 
We did not use continuum intensity maps in the image cross-correlation
because magnetic features were still observed after the following
sunspot became very small and the granulation patterns had changed 
significantly.
This allows us to trace the change of magnetic flux until the
disappearance of the sunspot. 

\subsection{Horizontal Velocity of Magnetic Elements}
Horizontal velocities of magnetic elements are necessary for estimation
of a flux transport rate.
Hereafter, the Na {\footnotesize D} line-of-sight magnetogram is defined as the
Stokes \textit{V} image divided by the simultaneous Stokes \textit{I} images. 
For each SP map (about 32 minutes), 16 line-of-sight magnetograms with
a 2 minute cadence were obtained. 
We made 8 horizontal velocity maps with a 4 minute cadence from the 16
line-of-sight magnetograms, using a local correlation tracking method
\citep[LCT;][]{November1988, Chae2001, Sakamoto2004}. 
The apodization window was a Gaussian with the FWHM of 1$\arcsec$ in the LCT. 
The LCT was applied for the pixels that have Stokes \textit{V/I} signals
larger than 0.0015 in two sequential line-of-sight magnetograms.
The threshold of 0.0015 was about 1$\sigma$ noise of the
line-of-sight magnetograms.
We assumed that the 1$\sigma$ noise of the line-of-sight magnetograms
corresponds to the width parameter (the standard deviation) of a
Gaussian fitted to a central part of a histogram for the signals in the
line-of-sight magnetogram (Fig.~\ref{flux_velocity}\textit{a}). 
When a pixel had a coss-correlation coefficient less than 0.9 or the
Stokes \textit{V/I} signal less than 0.0015, the horizontal velocity for
the pixel was assumed to be 0 km s$^{-1}$.

Finally, we averaged over the 8 horizontal velocity maps, and then
aligned the averaged map to the SP map obtained at the same period,
using the image cross-correlation between the line-of-sight magnetic flux
density with the SP and the line-of-sight magnetogram with the NFI.
Figure~\ref{flux_velocity}\textit{b} shows a histogram of the horizontal
velocity.
The average of the horizontal velocity is 0.5 km s$^{-1}$, which is similar
to the averaged speed of MMFs and moat flow in previous studies
\citep[e.g.][]{Brickhouse1988, Zhang2003}.  
Magnetic elements that move faster than the averaged moat flow
of 0.5-1.0 km s$^{-1}$ \citep{Hagenarr2005} are also detected. 
Furthermore, radial outward motions of magnetic elements can be seen
around the following sunspot in Figure~\ref{flux_velocity}\textit{c}.
Thus, radial outward motions of MMFs are successfully obtained
with the LCT.

\section{RESULTS}
\begin{figure}
\epsscale{1.0}
\plotone{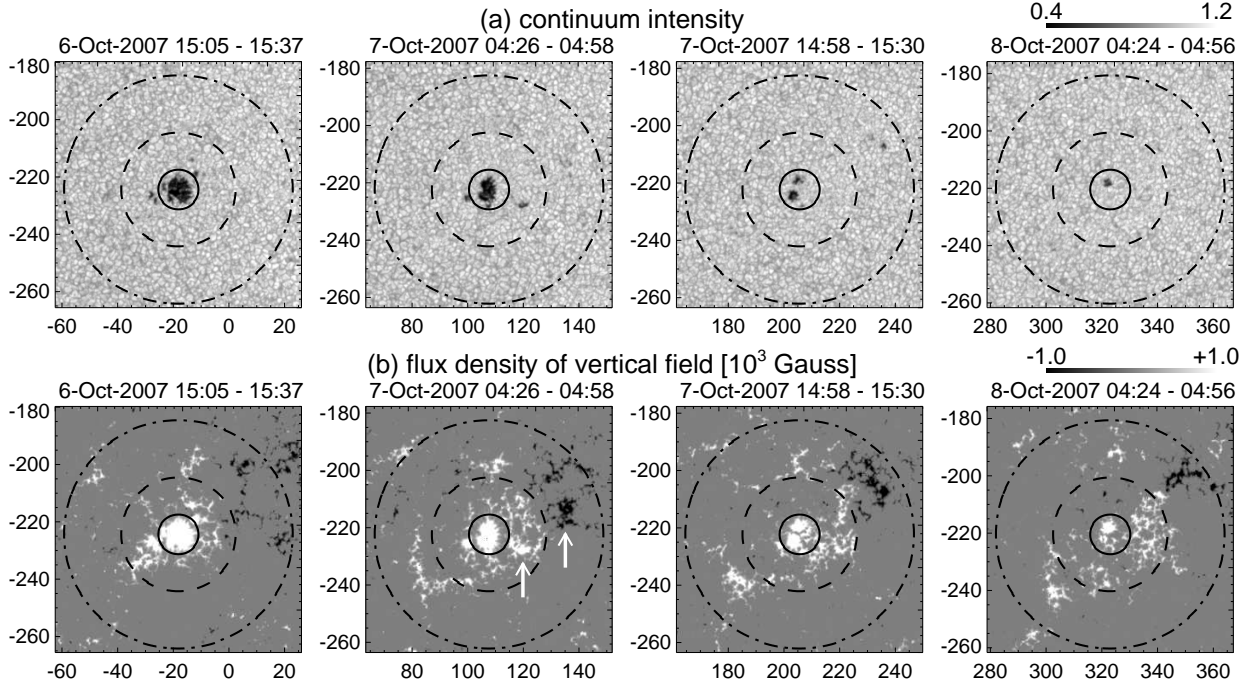}
\caption{Time series of (\textit{a}) continuum intensity and
 (\textit{b}) magnetic flux density of vertical field for the following
 sunspot in NOAA AR 10972. 
The solid, dashed, and dash-dotted circles indicate the outer boundaries
 of regions called the sunspot region (7$\arcsec$ from the sunspot
 center), the unipolar region (20$\arcsec$ from the sunspot center), and
 the mixed polarity  region (40$\arcsec$ from the sunspot center),
 respectively. 
The vertical and horizontal axes show the positions with respect to the
 disk center in units of arcseconds.
The white arrows indicate an emerging bipole.
}
\label{flux_image_evo}
\end{figure}

\begin{figure}
\epsscale{1.0}
\plotone{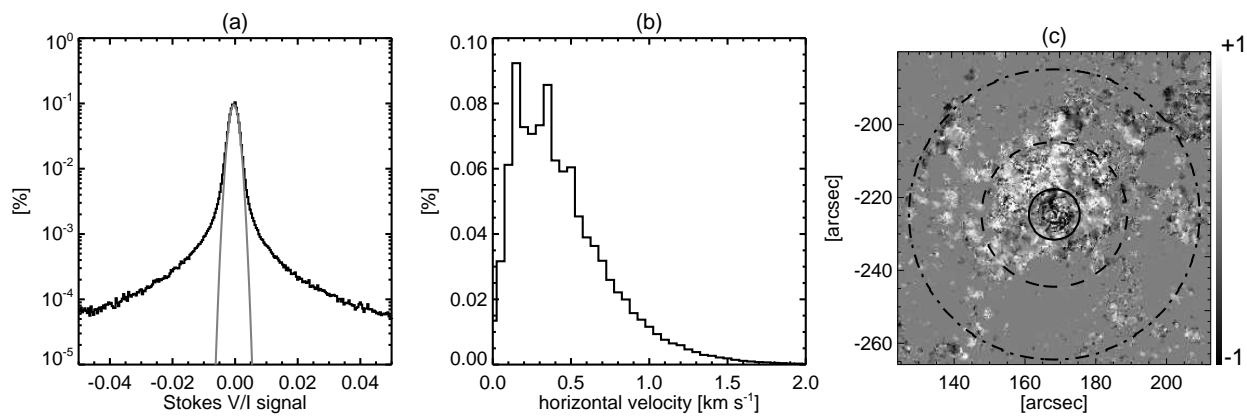}
\caption{(\textit{a}) Histogram of the magnetic signal in a line-of-sight
 NFI magnetogram (solid line).
The gray line shows a result of a Gaussian fitted to a central part of
 the histogram. 
(\textit{b}) Histogram of horizontal velocity calculated by the local
 correlation tracking technique that is applied for line-of-sight
 magnetograms. 
The horizontal velocities are averaged over the period taking each SP
 map (about 32 minutes).
(\textit{c}) Spatial distribution of a radial component of the horizontal
 velocity averaged over the period from 10:59:32 to 11:27:31 on 2007
 October 7 in units of km s$^{-1}$.
Positive corresponds to a radial outward motion.
The circles are same as those in Fig.~\ref{flux_image_evo}.
}  
\label{flux_velocity}
\end{figure}

Many moving magnetic features (MMFs) are observed outside the sunspot
(Fig.~\ref{flux_image_evo}\textit{b}).
The MMFs of positive polarity (the polarity of the following sunspot)
are dominant in the region with a radial distance ($r_s$) less than
20$\arcsec$ from the center of the following sunspot. 
MMFs of both polarities are located in the region $r_s > 20\arcsec$. 
Most magnetic elements with negative polarity are located in the
north-western side to the sunspot, and are in contact with positive
polarity elements.
Hereafter, we call regions with $r_s = 0\arcsec - 7\arcsec$,
$r_s = 7\arcsec - 20\arcsec$, and $r_s = 20\arcsec - 40\arcsec$ as the
sunspot region, the unipolar region, and the mixed polarity region,
respectively. 
In regular decaying sunspots, the unipolar region corresponds to the
inner and middle moat region, and the mixed polarity region corresponds
to the region around the outer boundary of the moat region.
We investigate the temporal change of magnetic flux in these three regions,
and also investigate how much magnetic flux passes the outer boundaries
of the three regions.

\begin{figure}
\epsscale{0.5}
\plotone{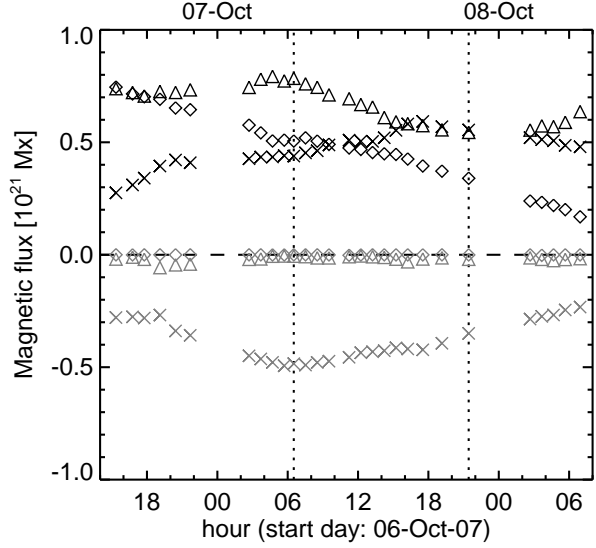}
\caption{Time profiles of total magnetic flux in the sunspot region
 (diamond), unipolar region (triangle), and mixed polarity region
 (cross) as determined from the SP maps. 
The black symbols show magnetic flux of positive polarity, and the
 gray symbols show magnetic flux of negative polarity.
The time profiles in the period between the two dotted lines are used for
 calculation of the flux change rate ($dF/dt$) in
 Fig.~\ref{flux_summary}. 
}
\label{flux_plot}
\end{figure}

Positive magnetic flux decreases at a constant rate in the sunspot
region, as shown in Figure~\ref{flux_plot}. 
In the unipolar region, the positive magnetic flux increases from 01:00
to 06:00 on October 7 as a result of the flux emergence indicated by the
arrows in Figure~\ref{flux_image_evo}.
The emerging bipoles appear close to $r_s = 20\arcsec$ and the opposite
polarities associated with the emergence separate in the the regions $r_s
< 20\arcsec$ ($r_s > 20\arcsec$) for positive (negative) flux.
This flux emergence also causes increase of the negative magnetic flux
in the mixed polarity region.
After the flux emergence, the positive flux in the unipolar region and
the negative flux in the mixed polarity region decrease at similar
rates. 
On the other hand, the positive flux in the mixed polarity region
increases during this period.
The negative magnetic flux in the sunspot and unipolar regions is much
smaller than the signed fluxes in the other regions.
We estimate an increase and decrease rate of magnetic flux in each
region by a linear fit to the time profiles from 06:15 to
21:26 on October 7 in Figure~\ref {flux_plot}. 
The magnetic flux changes nearly linearly during this period.
The result is shown as ``$dF/dt$'' in Figure~\ref{flux_summary}.
Even when the sunspot becomes too small on October 8, the positive flux
in the sunspot region still decreases at a constant rate.
On the other hand, the positive flux does not show any increases and
decreases in the unipolar region and mixed polarity region on October 8.
This tendency already may be seen in the later part on October 7.
The change of the tendency suggests the possibility of different flux loss
process in remnant active regions, but further, more detailed, investigations 
of this phenomena are needed.

\begin{figure}
\epsscale{1.0}
\plotone{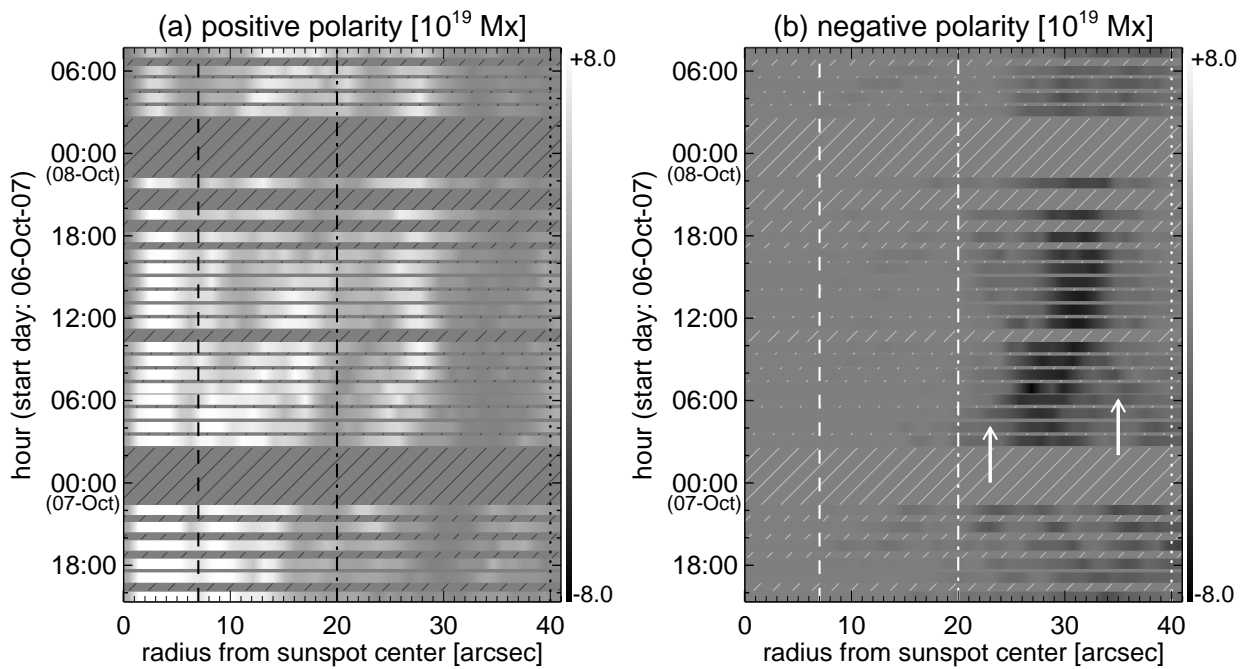}
\caption{Total magnetic flux at each radius from the
 sunspot center vs. time for (\textit{a}) positive
 polarity and (\textit{b}) negative polarity .
The total magnetic flux with positive (negative) polarity represents the
 sum total of magnetic flux for positive (negative) magnetic elements
 that have the same radius from the sunspot center. 
The hatched areas with oblique lines represent the periods without any
 SP observations. 
The dashed, dash-dotted, and dotted lines show the outer boundaries of the
 sunspot region (7$\arcsec$ from the sunspot center), the unipolar
 region (20$\arcsec$ from the sunspot center), and the mixed polarity
 region (40$\arcsec$ from the sunspot center), respectively. 
}
\label{flux_evo}
\end{figure}

Figure~\ref{flux_evo} shows temporal evolution of total magnetic flux at
each radius from the sunspot center.
The areas that have large positive magnetic flux move away from the sunspot
toward the mixed polarity region.
Such outward motion of the positive flux is due to magnetic flux
transported by the MMFs of positive polarity.
The outward motion of the positive flux stops around 30$\arcsec$ from
the sunspot center.
Figure~\ref{flux_image_evo} also shows that most of the positive
magnetic elements do not go out of the mixed polarity region, except for
the positive elements located in the north-eastern side to the sunspot.
Most of the negative magnetic flux is located around 30$\arcsec$
from the sunspot center, and converging motion of the negative flux is
observed around the 30$\arcsec$ from the sunspot center (indicated by
the arrows in Fig.~\ref{flux_evo}).
These results suggest that the boundary of a supergranular cell, which
corresponds to the outer boundary of the moat region, is located around
30$\arcsec$ from the sunspot center. 

We estimate a radial transport rate ($F_v(r_s)$) of magnetic flux at
each radius ($r_s$) from the sunspot center using the following formula:
\begin{equation}
F_v(r_s) = {\sum_{r_s=r_1}^{r_2}}\frac{|F(r_s)|}{\pi(r_2^2-r_1^2)}v_r(r_s)\frac{2\pi(r_1+r_2)}{2},\label{eq_flux_t}
\end{equation}
where $F(r_s)$ is the magnetic flux density of vertical fields
(Eq.~(\ref{eq_flux_d})) and $v_r(r_s)$ is the radial component of
horizontal velocity at $r_s$.
We set $r_1$ and $r_2$ as $r_s \pm 1\arcsec$, respectively.
The behavior of $F_v(r_s)$ with time in Figure~\ref{flux_transport}
shows that the motions of magnetic flux seen in Figure~\ref{flux_evo}
(e.g. outward motion of the positive flux, converging motion of the
negative flux) are obtained by using Equation~(\ref{eq_flux_t}).  
The inward flux transport rate for positive polarity elements around the
sunspot center is probably incorrect.
This anomaly is mainly due to the shrinkage of the sunspot, yielding
spurious inward $v_r(r_s)$:
the LCT does not work well when the contrast of line-of-sight
magnetograms is low.  
However, the flux transport within the sunspot region is not the object
of this study.
We focus on the flux transport rates at the outer edges of the sunspot
region as well as the unipolar and mixed polarity regions
during the period when the flux change rate is calculated from
Figure~\ref{flux_plot} (06:15 to 21:26 on October 7).
The flux transport rates averaged over this period are summarized as
``$F_v$'' in Figure~\ref{flux_summary}. 
The direction of the arrows shows whether magnetic flux is transported
radially inward or outward.

\section{DISCUSSION AND CONCLUSIONS}
\begin{figure}
\epsscale{1.0}
\plotone{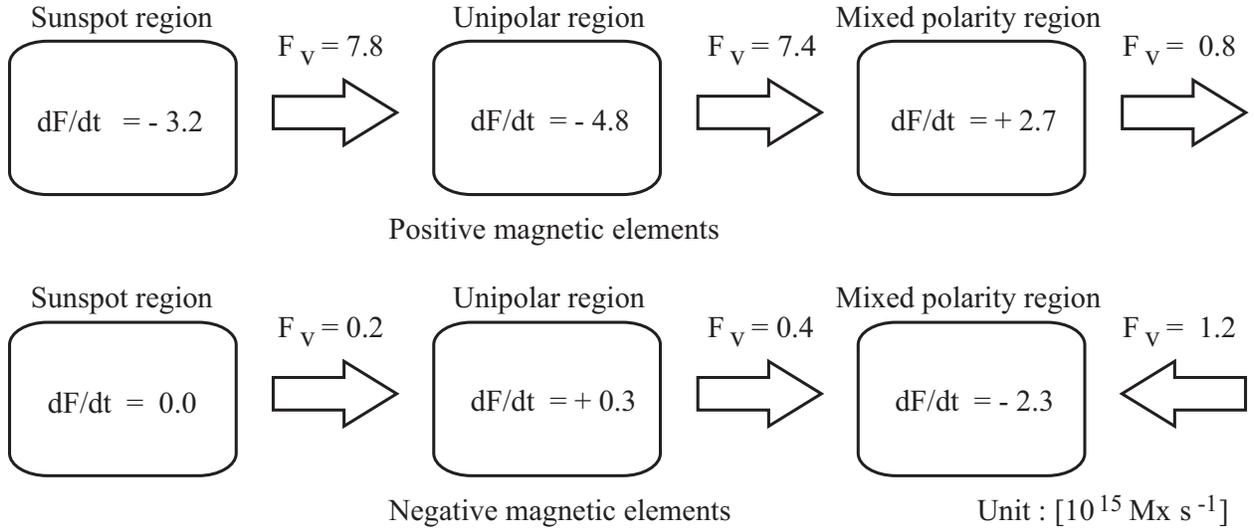}
\caption{Summary of observations for the flux change rate ($dF/dt$) and
 the flux transport rate ($F_v$) in units of $10^{15}$Mx s$^{-1}$.}
\label{flux_summary}
\end{figure}

\begin{figure}
\epsscale{1.0}
\plotone{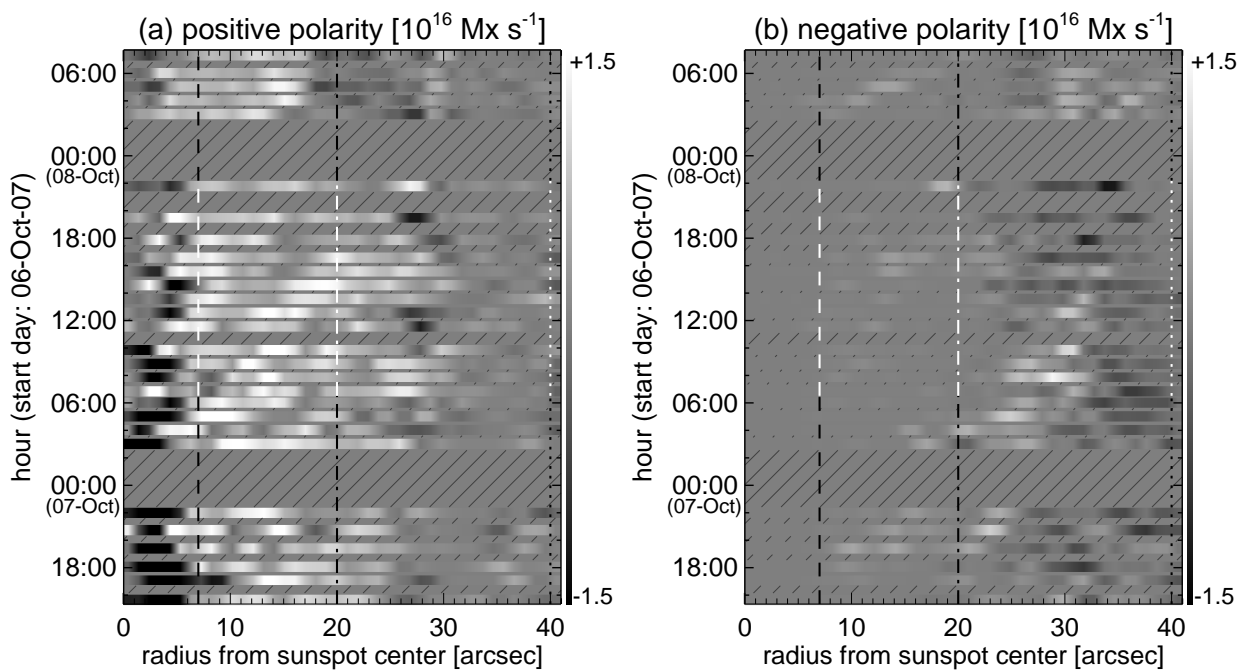}
\caption{Same as Fig.~\ref{flux_evo}, but for showing the flux
 transport rate vs. time. 
The flux transport rate ($F_v(r_s)$) is calculated as
 Eq.~(\ref{eq_flux_t}).
The white indicates magnetic flux transported away from the sunspot center. 
The value averaged along the white part of each line is identical to
 the flux transport rate ($F_v$) in Fig.~\ref{flux_summary}. 
}
\label{flux_transport}
\end{figure}

The averaged flux change and averaged flux transport in
Figure~\ref{flux_summary} are described in units of Mx s$^{-1}$, so
that we can compare these values directly. 
The observed flux change [$(dF/dt)_{Obs}$] in each region and the
observed flux transport [$(F_v)_{Obs}$] at its boundaries would have a
relationship:
\begin{equation}
(\frac{dF}{dt})_{Obs} = (\frac{dF}{dt})_{Emerge} -
 (\frac{dF}{dt})_{Loss} \pm (F_v)_{Obs}.\label{eq_flux_relation}
\end{equation}
The increase of magnetic flux due to flux emergence
[$(dF/dt)_{Emerge}$] can be neglected in this case, because we select
the period without any significant flux emergence.
Thus, we can estimate an actual flux loss rate [$(dF/dt)_{Loss}$] in each
region from the observations. 

The total of flux decrease rates in the sunspot and unipolar regions
($dF/dt$ = -3.2 - 4.8 = -8.0$\times 10^{15}$ Mx s$^{-1}$) is almost equal to
the flux transport rate at the outer boundary of the unipolar region for
the positive polarity ($F_v$ = 7.4$\times 10^{15}$ Mx s$^{-1}$). 
This means that most of magnetic flux that disappeared in the sunspot
and unipolar regions is carried away to the mixed polarity region.
Note that we do not trace each magnetic element, and all of the
MMFs that separated from the sunspot may not reach at the outer
boundary of the unipolar region.
However, we make Figure~\ref{flux_summary} from the observations for about
12 hr, which is twice as long as the period that magnetic elements
with the average horizontal speed of 0.5 km s$^{-1}$ need to move through
the unipolar region of a width of 14$\arcsec$. 
The increase of the positive flux in the mixed polarity region supports the
migration of positive flux into the mixed polarity region.
Both the increase of the positive flux in the mixed polarity region and
the flux transport for the positive polarity elements at the outer
boundary of the mixed polarity region are smaller than the positive flux
transported from the unipolar region. 
Therefore, the magnetic flux that is carried away from the sunspot (and moat
region) mostly disappears in the mixed polarity region, especially near
the outer boundary of the moat region.

One issue is that the positive flux carried away from the sunspot
region ($F_v$ = 7.8$\times 10^{15}$ Mx s$^{-1}$) is bigger than decrease
of the positive flux in the sunspot region ($dF/dt$ = -3.2$\times
10^{15}$ Mx s$^{-1}$). 
This tendency was also reported in the previous work with a lower
(about $1\arcsec$) spatial resolution \citep{Kubo2007a}.
As a result of no flux emergence in the sunspot region,
the flux transport rate should be less than the flux decrease
rate in the sunspot region.
That is to say that the flux transport rate is overestimated at the
outer boundary of the sunspot region. 
In the calculation of horizontal velocities, we use the apodization
window with 1$\arcsec$, which is lower than the spatial resolution of
the magnetic field maps.
Such a lower spatial resolution of the horizontal velocity maps probably
causes the overestimation of the flux transport rate at the outer
boundary of the sunspot region.
Fuzzy, small magnetic elements with a short lifetime have been observed
around the outer boundary of decaying sunspots \citep{Zhang2007, Kubo2008}.
These fuzzy magnetic elements have higher outward motion and smaller
magnetic flux than those of usual MMFs.  
In the estimation of flux transport at the outer boundary of the sunspot,
magnetic flux is mostly represented by MMFs with small horizontal
velocity and large magnetic flux, but its horizontal velocity is
represented by the fuzzy magnetic elements.
We believe that these fuzzy magnetic elements correspond to a
fluctuation of field strength or a fluctuation of inclination of penumbral
magnetic fields, and thus do not contribute to the flux loss of the sunspot. 
Further investigation using spectropolarimetric measurements with
a higher cadence will be necessary to know the nature of such fuzzy
magnetic elements and their impact on the presented calculations.

The magnetic flux of negative polarity decreases in the mixed polarity
region, although the negative flux converges from the inner and outer
boundaries of the mixed polarity region.
Considering that the negative flux moves into the mixed polarity region
with the average rate of 1.6$\times 10^{15}$ Mx s$^{-1}$, the actual flux
loss rate [$(dF/dt)_{Loss}$] in the mixed polarity region may be as
large as 3.9$\times 10^{15}$ Mx s$^{-1}$ from
Equation~(\ref{eq_flux_relation}).  
This flux loss rate of negative polarity is balanced by the actual flux
loss rate of the positive polarity (3.9$\times 10^{15}$ Mx s$^{-1}$), 
which is a difference between the flux transport rate ($F_v$ = 7.4 - 0.8
= 6.6$\times 10^{15}$ Mx s$^{-1}$) into the mixed polarity region and the
flux increase rate ($dF/dt$ = 2.7$\times 10^{15}$ Mx s$^{-1}$) there.
The flux loss rates with both polarities in the mixed polarity region are
consistent with the cancellation rates in active regions
\citep{Chae2000,Chae2004,Kubo2007b}.  
Furthermore, most of the magnetic elements with negative polarity are
located in contact with the positive elements.
These results suggest that magnetic flux cancellation at the outer
boundary of the moat region is essential for the removal of the sunspot
magnetic flux from the photosphere.

\acknowledgments
We would like to thank S. Tsuneta, Y. Katsukawa, T. Yokoyama,
and S. W. McIntosh for useful discussions and comments on this paper. 
We also acknowledge Y. Sakamoto for development of the LCT programs.
\textit{Hinode} is a Japanese mission developed and launched by
ISAS/JAXA, with NAOJ as domestic partner and NASA and STFC (UK) as
international partners. It is operated by these agencies in cooperation
with ESA and NSC (Norway). 
This work was partly carried out at the NAOJ \textit{Hinode} Science Center,
which is supported by the Grant-in-Aid for Creative Scientific Research
``The Basic Study of Space Weather Prediction'' from MEXT, Japan (Head
Investigator: K. Shibata), generous donations from Sun Microsystems, and
NAOJ internal funding.


\end{document}